\documentclass[12pt]{article}
\usepackage{amssymb,latexsym}
\usepackage[mathscr]{eucal}
\usepackage{pstricks,pst-node,pst-text,pst-3d}
\newcommand{\ft}[2]{{\textstyle\frac{#1}{#2}}}

\newcommand{\Z}{{\mathbb Z}}
\newcommand{\R}{{\mathbb R}}
\newcommand{\C}{{\mathbb C}}

\newcommand{\beq}{\begin{equation}}
\newcommand{\eeq}{\end{equation}}
\newcommand{\beqa}{\begin{eqnarray}}
\newcommand{\eeqa}{\end{eqnarray}} 
 \newcommand{\uno}{\mbox{1
\kern-.59em {\rm l}}}

\newcommand{\tinyyoung}[1]{\mbox{\tiny\young(#1)}}
\usepackage[vcentermath,enableskew]{youngtab}
\newcommand{\nn}{\nonumber}
\begin{document}
%%%%%%%%%%%%%%%%%%%%%%%  FRONTESPIZIO  %%%%%%%%%%%%%%%%%%%%%%%%%%%%%%%%%%%
\begin{titlepage}
\begin{flushright}
{ROM2F/2004/21}\\
\end{flushright}
%%%%%%%%%%%%%%%%%%%%%%%  TITOLO  %%%%%%%%%%%%%%%%%%%%%%%%%%%%%%%%%%%%%%%%%
\begin{center}
{\large \sc Instantons on Quivers and Orientifolds}\\

\vspace{0.2cm}
%%%%%%%%%%%%%%%%%%%%%%%%  AUTORI  %%%%%%%%%%%%%%%%%%%%%%%%%%%%%%%%%%%%%%%%%
{\sc Francesco Fucito}\\
{\sl Dipartimento di Fisica, Universit\'a di Roma ``Tor Vergata'',
I.N.F.N. Sezione di Roma II\\
Via della Ricerca Scientifica, 00133 Roma, Italy}\\
{\sc  Jose F. Morales}\\
{\sl Laboratori Nazionali di Frascati \\
P.O. Box, 00044 Frascati, Italy }
and\\
{\sc Rubik Poghossian}\\
{\sl Yerevan Physics Institute\\
Alikhanian Br. st. 2, 375036 Yerevan, Armenia}\\
\end{center}
\vskip 0.5cm
%%%%%%%%%%%%%%%%%%%%%%%%  ABSTRACT  %%%%%%%%%%%%%%%%%%%%%%%%%%%%%%%%%%%%%%
\begin{center}
{\large \bf Abstract}
\end{center}
{We compute the prepotential for gauge theories descending from ${\cal
N}=4$ SYM via quiver projections and mass deformations.
 This accounts for gauge theories with product gauge groups and
 bifundamental matter.  The case of
massive orientifold gauge theories with gauge group SO/Sp is also
described. In the case with no gravitational corrections the
results are shown to be in agreement with Seiberg-Witten analysis
and previous results in the literature.
%A self-consistent review of the localization algorithm is included.
}

\par    \vfill
\end{titlepage}
%%%%%%%%%%%%%%%%%%%%%%%%%%%%%%%%%%%%%%%%%%%%%%%%%%%%%%%%%%%%%%%%%%%%%%%%%%
%%%%%%%%%%%%%%%%%%%%%%  INIZIO TESTO  %%%%%%%%%%%%%%%%%%%%%%%%%%%%%%%%%%%%
%%%%%%%%%%%%%%%%%%%%%%%%%%%%%%%%%%%%%%%%%%%%%%%%%%%%%%%%%%%%%%%%%%%%%%%%%%
\addtolength{\baselineskip}{0.3\baselineskip}
%%%%%%%%%%%%%%%%%%%%%%%%%%%%%%%%%%%%%%%%%%%%%%%%%%%%%%%%%%%%%%%%%%%%%%%%%%
%%%%%%%%%%%%%%%%%%%%%%%%%%%%%%%%%%%%%%%%%%%%%%%%%%%%%%%%%%%%%%%%%%%%%%%%%%
%%%%%%%%%%%%%%%%%%%%%%%%%%%%%%%%%%%%%%%%%%%%%%%%%%%%%%%%%%%%%%%%%%%%%%%%%%
%%%%%%%%%%%%%%%%%%%%%%%%%%%%%%%%%%%%%%%%%%%%%%%%%%%%%%%%%%%%%%%%%%%%%%%%%%
%%%%%%%%%%%%%%%%%%%%%%  SECTION 1     %%%%%%%%%%%%%%%%%%%%%%%%%%%%%%%%%%%%
%%%%%%%%%%%%%%%%%%%%%%%%%%%%%%%%%%%%%%%%%%%%%%%%%%%%%%%%%%%%%%%%%%%%%%%%%%
\setcounter{section}{0}
\section{Introduction}
Localization techniques have proved to be a very powerful tool in
the analysis of supersymmetric gauge theories (SYM). For some
years there was a very slow progress in multi-instanton
computations  due to the difficulty of properly treating the ADHM
constraints (see \cite{Dorey:2002ik} for a review and a complete
list of references). Early attempts to introduce localization were
not very fruitful \cite{Bellisai:2000bc,Hollowood:2002ds} due to
the non-compactness of the moduli space of gauge connections
\footnote{For a comparison with analogous problems for the moduli spaces
of punctured Riemann surfaces see \cite{Bertoldi:2004cc}.}. At
last a proper treatment of these problems was suggested
\cite{Nekrasov:2002qd} (see also
\cite{Flume:2002az,Bruzzo:2002xf}). The key new ingredient is the
introduction of a deformation (by turning gravitational
backgrounds) of the theory which localizes the ADHM integrals
around certain critical point of a suitable defined equivariant
differential. The ADHM construction of self-dual connections in
the presence of gravitational backgrounds have been worked it out
in \cite{Flume:2004rp}.

The computation of the partition function for a SYM theory can
then be carried out by expanding around  critical points and
keeping only quadratic terms in the fluctuations. The final result
is the inverse of the determinant of the Hessian. The computation
of the eigenvalues of this determinant becomes now the real
problem. The symmetries which deform the theory to study the
equivariant cohomology come now to the rescue. Since the saddle
points are critical points for these symmetries, the tangent space
can be decomposed in terms of the eigenvalues with respect to the
deforming symmetries. This allows the computation of the character
and, as a consequence, of the eigenvalues of the Hessian
\cite{nakajima}. This computation was carried out for ${\cal N}=2,
2^*, 4$ theories in \cite{Flume:2002az,Bruzzo:2002xf}. See
\cite{Bruzzo:2003rw} for a treatment of the localization formula
on a supermanifold.

 The character is an extremely versatile
function of the eigenvalues. Further symmetries of the theory can
be incorporated into it  with no need of further computation. A
demonstration of this statement is given in this paper in which we
compute the partition function of some quiver gauge theories and
in a companion paper in which we compute the partition functions
of  ${\cal N}=2, 2^*, 4$ SYM theories on ALE manifolds\cite{fmp}.

We have now to specify which symmetries we are referring to. The moduli spaces of gauge connections
for the ${\cal N}=4$ SYM naturally arise in the D(-1)-D3-brane system \cite{wttn,Douglas:1998uz}.
The presence of the D3-brane naturally splits the $SO(10)$, originally acting on the ten dimensional space,
into $SO(4)\times SO(6)$. $SO(6)$ is the internal symmetry of the ${\cal N}=4$ SYM. To get a theory
with a smaller number of supersymmetries we can mod the compactified dimensions by a discrete group
$\Z_p$ \cite{Hollowood:1999bm,Fucito:2001ha}.
Modding instead the space-time by the discrete group we find ALE manifolds \cite{fmp}. The crucial point
is that these discrete groups act trivially on the fixed points of the deformed theory. This implies that the
new results can be extracted from the $\R^4$ computations of \cite{Flume:2002az,Bruzzo:2002xf}
by doing proper projections on the tangent spaces.

In this paper we try to present these facts without referring to
the formalism of the ADHM construction of gauge connections which
is the standard point of view that can be found in literature.
Hoping to make our results clearer to the string-oriented reader,
our point of view will be strictly the one of the system of $k$
D(-1) and $N$ D3-branes. In section 2 we describe the massless
content of this system and derive the BRST transformations on the
moduli space from SUSY transformations in two dimensions (that we
discuss in appendix A for completeness). We then discuss the
computations of the integral on the moduli space and of the
character. In the first part of Section 3 we apply an orbifold
projection $\Gamma=\Z_p$ on the moduli space and derive the
corresponding character. This accounts for quiver theories
\footnote{See \cite{halpern} for an early discussion on quivers.} with
product gauge group $\prod_{q=0}^{p-1} U(N_q)$ and matter in the
bifundamental representations $(N_q,\bar N_{q+1}), (N_{q+1},\bar
N_q)$. The second part deals with explicit computations. We are
able to treat a wide range of models: massive or massless with
various matter content. The largest formulae are confined
in appendices B, C. Appendix B is a check against existing
literature
\cite{Ennes:1998ve,Feichtinger:1999vt,Gomez-Reino:2003cp}. Finally
in Section 4, instead of modding by a discrete group, we introduce
an orientifold plane in the D(-1)-D3-brane system. The resulting
SYM has $Sp$ or $SO$ gauge group and has been recently studied in
\cite{Marino:2004cn,Nekrasov:2004vw}. The character for this
system quickly leads to the results for the cases with mass in the
adjoint and fundamental representations. In addition, the
character formulae derived here compute directly the determinant
at the fixed point as a product over $4kN$ (the dimension of the
moduli space) eigenvalues in contrast with the approach followed
in \cite{Marino:2004cn,Nekrasov:2004vw} where the same determinant
is expressed in terms of a residue of a ratio of $4kN+4k^2$ and
$4k^2$ eigenvalues.
 We believe that this simplification can be useful for future developments.

\section{D-instantons}
\setcounter{equation}{0}
\subsection{ADHM  manifold}

 The moduli space of self-dual solutions of $U(N)$ YM-equations
in four dimensions is elegantly described by a real $4kN$-dimensional
hypersurface embedded in a $4kN+4k^2$-dimensional ambient space
via the ADHM constraints. Alternatively one can think of the ADHM
constraints as the D and F-flatness conditions of an auxiliary gauge
theory of $U(k)$ matrices. The relevant theory is the $U(k)$ gauge
theory describing the low energy dynamics of  a stack of $k$ D(-1)-branes
superposed to a stack of $N$ D3-branes. D3-brane descriptions of
four dimensional gauge theories are by now of common use. In this
formalism a $U(N)$ gauge theory is realized by the lowest modes of
the open strings ending on a stack of $N$ D3-branes. These include
an adjoint vector field and adjoint
scalars whose $N$ eigenvalues parameterize the positions of the
$D3$ branes in the transverse space. The number of flat transverse
directions can be chosen to be $ 2, 6$ according to whether one
considers an ${\cal N}=2,4$ gauge theory. The ${\cal R}$-symmetry
groups $SU(2), SO(6)$ act as rotations on this internal
space. D(-1)-branes are to be thought of as instantons in the D3-brane gauge theory
with multi-instanton moduli now associated to open strings
ending on the D(-1)-instanton stack.

We start by considering the ${\cal N}=4$ supersymmetric gauge theory. The
${\cal N}=4$ SYM lagrangian follows from the dimensional reduction of the ${\cal N}=1$ SYM in
$D=10$ down to the $D=4$ worldvolume of the D3-branes.
The presence of the D3-branes breaks the Lorentz symmetry of the
ten-dimensional space seen by the D(-1)-instantons down to
$SO(4)\times SO(6)_{\cal R}$. The resulting gauge
theory living on the D(-1)-instanton worldvolume follows
from the dimensional reduction
down to zero dimension of a ${\cal N}=1$ $U(k)$ SYM theory in
$D=6$ with an adjoint (D(-1)-D(-1) open strings) and $N$
fundamental (D(-1)-D3 open strings) hypermultiplets. In the
notation of ${\cal N}=1$ in $D=4$ this corresponds to a $U(k)$ vector multiplet,
three adjoint chiral multiplets and $2N$ fundamental chiral multiplets.

 For our purposes
 it is convenient to view the above multiplets from a $D=2$ perspective.
 The two-dimensional plane where the $D=2$ gauge theory lives will be parameterized by a complex scalar field $\phi$.
 The choice of the plane is a matter of convention, it corresponds to choose a
${\cal N}=2$ subalgebra inside ${\cal N}=4$.
 The field $\phi$ plays the role of a connection in the resulting $D=2$ gauge theory and
 it will be crucial in the discussion of localization below.
  Localization is based on the existence of a  BRST current $Q$
\cite{Dorey:2000zq,Bellisai:2000bc,Fucito:2001ha} in the moduli
space. The charge $Q$ is BRST in the sense that it squares to zero up
to a symmetry transformation parameterized by $\phi$.
 Deformations of the gauge theory like vevs or gravitational backgrounds will
 correspond to distribute the $k$ D(-1)-instantons along the $\phi$-plane.

 Let us start first by describing the field content of the
 theory. For convenience we use the more familiar $D=4$ ${\cal N}=1$
 notation to describe the $D=6$ starting gauge theory.
 The starting point is a ${\cal N}=1$ $U(k)$ gauge theory
 in $D=4$ with vector multiplet $V=(\phi,B_4| \eta,\chi_{\R},{\cal
 M}_4|H_{\R} )$, three adjoint chiral multiplets
 $C_{\ell=1,2,3}=(B_{\ell}|{\cal M}_{\ell},\chi_{\ell 4}|H_{\ell 4})$
and $2N$ chiral multiplets in the fundamental
$C_{\dot{\alpha}}=(w_{\dot{\alpha}}|\mu_{\dot{\alpha}},\mu_{\dot{a}}|H_{\dot{a}})$,
$\dot{\alpha}=1,2$, $\dot{a}=2+\dot{\alpha}$. Here all fields are
complex except $\chi_{\R},\eta,H_{\R}$.
 In particular the
five complex scalars describing the instanton position in the
ten-dimensional space are denoted by $\phi,B_\ell$ \footnote{In
the $D=6$ notation the six-dimensional vector $\chi_a$ is built
out of the three complex scalars $\phi,B_{3,4}$. See
\cite{Dorey:2002ik} for more details.}. $\phi$ and $B_4$ describe the
four real components of the vector field, while
$w_{\dot{\alpha}}=(I,J^\dagger)$ describe D(-1)-D3 string
modes\footnote{$(I,J^\dagger)$ are $k\times N$ matrices which
constitute the first $N$ rows in the ADHM matrix. In the ${\cal
N}=4$ theory, we are currently describing, the usual ADHM
construction needs to be generalized. See
\cite{Dorey:2000zq,Fucito:2001ha} for a discussion of this
point.}. Fermions and auxiliary fields (here denoted by $H$)
complete the supermultiplets.
The BRST transformations of these fields are obtained in two steps.
First we take the  variations of ${\cal N}=1$
supersymmetry in $D=4$ and compactify to $D=2$ \cite{Witten:1993yc}.
The resulting theory has ${\cal N}=(2,2)$ supersymmetry.
The content of the ${\cal N}=(2,2)$ vector
and chiral multiplets, their SUSY transformations and the translation to instanton variables
are given in Appendix \ref{ad2}.
Second the BRST charge, $Q$, is defined by
choosing $\bar{\epsilon}_-=\epsilon_-=0$,
$\bar{\epsilon}_+=-\epsilon_+=i\xi$ in (\ref{mult1},\ref{mult2})
\footnote{SUSY and BRST transformations ($\delta$ and $Q$ respectively) are put in relation
choosing $\delta=\xi Q$.
%$\xi$ is the parameter of the SUSY transformation.
}
and further compactifying to zero dimension.
The covariant derivatives $-4i \nabla_{--}$ are then replaced by the
connection $\phi$. From (\ref{mult2p}) one finds
 \beqa\label{brs}
Q \,B_\ell&=&{\cal M}_\ell \ \;\qquad
Q\, {\cal M}_\ell= [\phi,B_\ell]  \nonumber\\
Q \,w_{\dot{\alpha}}&=& \mu_{\dot{\alpha}} \;\qquad ~~Q \,\mu_{\dot{\alpha}}=\phi\, w_{\dot{\alpha}}\nonumber\\
Q\, \mu_{\dot{a}}&=& H_{\dot{a}} \ \;\qquad Q\, H_{\dot{a}}= \phi\, \mu_{\dot{a}}\nonumber\\
Q \,\chi_{[\ell_1\ell_2]}&=& H_{[\ell_1 \ell_2]} \ \;\quad ~Q \,H_{[\ell_1 \ell_2]}= [\phi,\chi_{[\ell_1\ell_2]}]\nonumber\\
Q \,\chi_{\R}&=& H_{\R} \ \;\quad ~~~~~Q \,H_{\R}= [\phi,\chi_{\R}]\nonumber\\
Q \,\bar{\phi}&=& \eta  \;\qquad ~~~Q\, {\eta}= [\phi,\bar{\phi}]\nn\\
 Q\,\phi&=&0\ \ , \eeqa
  Here $\ell_{1,2}=1,\ldots 4$ and $\chi_{[\ell_1 \ell_2]}$,$H_{[\ell_1 \ell_2]}$ are antisymmetric
  two-tensors with six real components.

  The $H$ are auxiliary fields (D or F terms) implementing the generalized ADHM constraints
 \beqa\label{mommap4}
{\cal E}_{\R}&=&[B_\ell,B_\ell^\dagger]+II^\dagger-J^\dagger J -\zeta=0 \nonumber\\
{\cal E}_{12}&=&[B_1,B_2]+[B_3^\dagger, B_4^\dagger]+IJ=0\nn\\
{\cal E}_{13}&=&[B_1,B_3]-[B_2^\dagger, B_4^\dagger]=0\nn\\
{\cal E}_{14}&=&[B_1,B_4]+[B_2^\dagger, B_3^\dagger]=0 \nn\\
{\cal E}_{1}&=&B_3 I-B_4^\dagger J^\dagger=0 \nn\\
{\cal E}_{2}&=&B_4 I+B_3^\dagger J^\dagger=0 \eeqa with ${\cal
E}_{AB}\equiv \ft12 \epsilon_{ABCD} {\cal E}^\dagger_{CD}$.
 It is convenient to collect all bosonic (except $\phi$) and fermionic fields in the l.h.s.
 of (\ref{brs}) into two vectors: the (super)moduli are those that appear in the l.h.s.
of (\ref{brs}), while the Q-differentials or tangent vectors are those in the r.h.s.
\beqa
&& \vec{m}=(B_\ell,w_{\dot{\alpha}})  \qquad \vec\Psi= Q\,\vec m \nn\\
&& \vec\chi=(\chi_{[\ell_1\ell_2]},\mu_{\dot{a}}) \qquad \vec
H=Q\, \vec\chi \nn
\label{mod+tang}
\eeqa with $Q^2 A=\phi\cdot A$
where $\phi\cdot A$ is given by $[ \phi,A]$,
  $\phi A$ or $A \phi$ according to whether $A$ transform in the adjoint, fundamental or antifundamental
  representation of $U(k)$.

 In terms of these variables the multi-instanton action
reads \cite{Fucito:2001ha,Dorey:2000zq}
\beq S= Q\, {\rm Tr}
\left[  \ft14 \,\eta [\phi,\bar{\phi}]+\ft12\, {\cal E}_{\R}\,
\chi_{\R}+ \vec{\cal E} \,\vec{\chi}+ \vec{H} \, \vec\chi +
\vec{m}\,(\bar{\phi}\cdot \vec\Psi)  \right]
\label{N=4}
\eeq
The
convention for the vector product is $\vec{\chi}\,\vec{H}\equiv
\sum_{i<j}\, \chi_{[\ell_i \ell_j]}^\dagger H_{[\ell_i \ell_j]}$.

Less supersymmetric multi-instanton actions are then found via
suitable orbifold projections or mass deformations of the original
${\cal N}=4$ theory. Additional ${\cal N}=2$ fundamental matter can be added
by introducing D7-branes. This leads to extra instanton moduli
arising from D(-1)-D7 strings and transforming as
\beq
 Q\,{\cal K}_f = H_f \qquad  Q\,H_f = \phi\,  {\cal K}_f
\eeq

\subsection{D-instanton partition}

The next step is to evaluate the centered D-instanton partition
function
\beq{\cal Z}_k = {1\over {\rm vol}U(k)}\int d\phi\,
 d\vec{m}\,d\vec{H}\, d\vec{\Psi}\,
d\vec{\chi}\,(d\bar{\phi}\,\,dH_{\R} d\chi_{\R}\, d\eta)\,
e^{-S}=\int_M  \, e^{-S} \label{int} \eeq The integral (\ref{int})
can be performed by means of the localization formula
\cite{Bruzzo:2002xf,Bruzzo:2003rw} \beq \int_M
e^{-S}=\sum_{\phi_0}{ e^{-S_0} \over {\rm Sdet}^{1\over 2}\, {\cal
L}_{\phi_0}} \label{locth} \eeq where the sum is running over the
fixed points $\phi_0$ (here assumed to be isolated). The operator
in the denominator of (\ref{locth}) is defined starting from the
super vector field  $Q$ (the BRST charge) on the ADHM manifold with
${\cal L}_{\phi_0}v=[Q,v]\vert_{\phi_0}$ on  a generic element $v$ of the
tangent space.

That the integral (\ref{int}) localizes is not surprising since
it computes an "index" of a supersymmetric theory. The crucial
observation in \cite{Nekrasov:2002qd} is that after suitable
 deformations (vev and gravitational
backgrounds)it localizes on a finite point
set $\{ \phi_0 \}$. The action, $S$, can be deformed using the
symmetries of the ADHM manifold. In fact, the symmetries of the
action (\ref{N=4}) are those that leave the constraints
(\ref{mommap4}) invariant. They are given by $g=U(k)\times
U(N)\times SO(4)_{\epsilon_{1,2}}\times
SO(4)_{\epsilon_{3,4}}$ with $SO(4)_{\epsilon_{3,4}}\in
SO(6)_{\cal R}$ the ${\cal R}$-symmetry subgroup preserved by the
choice of $Q$.  In particular the quartet indicated between
parenthesis in (\ref{int}) is a singlet of $g$ and therefore its
bosonic and fermionic contributions to the superdeterminant cancel
against each other and can be discarded.

The $U(k)$ invariance can be fixed by choosing $\phi$ along the
$U(1)^k$ Cartan subgroup
 $$
 \phi=(\phi(1),\phi(2),\ldots \phi(k))
 $$
 This leads to an additional Jacobian factor $\prod_{s<s'} (\phi(s)-\phi(s^\prime))$:
the Vandermonde determinant.
The determinant can be exponentiated by introducing an additional
BRST pair of auxiliary fields \beq
 Q \chi_{\C} =H_{\C} \qquad    Q H_{\C} =[\phi,\chi_{\C}]
\label{newmod+tang} \eeq The old set of auxiliary fields in
(\ref{mod+tang}) can be extended to include those in
(\ref{newmod+tang}). We denote the new set by a hat \beqa
&&\hat{\vec\chi}=(\chi_{\C},\chi_{[\ell_1\ell_2]},\mu_{\dot{a}})
\qquad \hat{H}=Q\, \hat{\vec\chi} \label{chin} \eeqa
The action and D-instanton partition function are
again given by (\ref{N=4}) and (\ref{int}) but now in terms of hatted fields with
${\cal E}_{\C}=0$. Indeed the action is quadratic in
$\chi_{\C},H_{\C}$ and after integrating them out the Vandermonde
determinant is reproduced. Notice that after the inclusion of
$\chi_{\C}$, the number of fermionic degrees of freedom in
$\hat{\vec \chi}$ matches that of  the bosonic ones in $\vec{m}$.

The BRST transformations (\ref{brs}) can be deformed to take into
account the action of $g$. For the purposes of the present paper
we will consider the symmetries given by the elements in the
Cartan of  $g$ i.e. $g\in U(1)^k \times U(1)^N\times
U(1)^2_{\epsilon_{1,2}}\times U(1)^2_{\epsilon_{3,4}}$
with $g$ parameterized by
$\phi(s),a_\alpha,\epsilon_\ell$ with $s=1,\ldots k$,
$\alpha=1,\ldots N$, $\ell=1,\ldots 4$. To this
end it is convenient to introduce the auxiliary $k$ and $N$ dimensional spaces
$V,W,$ transforming in the fundamental of $U(k)$, $U(N)$ and four
one dimensional spaces $Q_\ell$ carrying $U(1)_\ell$ charges.
\beqa && g\, Q_\ell=T_\ell \,Q_\ell \qquad T_\ell= e^{i
\epsilon_\ell} \qquad
\epsilon_\ell=(\epsilon_1,\epsilon_2,m,-m-\epsilon)\nn\\
&& g\, V=T_\phi V  \qquad   T_\phi=e^{i \phi} \nn\\
&& g\, W=T_a W  \qquad   T_{a}=e^{i a}  \eeqa with
$\epsilon=\epsilon_1+\epsilon_2$. The deformed BRST
transformations now squares to the infinitesimal
transformation  $\delta_{\phi,a,\epsilon_\ell}\in g$
  given by
\beqa
&&\delta_{\phi,a,\epsilon_\ell} Q_\ell=\epsilon_\ell \,Q_\ell
\qquad \epsilon_\ell=(\epsilon_1,\epsilon_2,m,-m-\epsilon)\nn\\
&&\delta_{\phi,a,\epsilon_\ell}\, V= [\phi, V]  \qquad
\delta_{\phi,a,\epsilon_\ell}\, W= a W \eeqa
The
transformations properties of the fields with respect to $g$ can then be
compactly summarized by \beqa
  B_\ell :&& V^*\times V\times (Q_\epsilon+Q_m)    \nn\\
\bar{B}_\ell: &&  V^*\times V\times \left[\Lambda^2 Q_\epsilon\times  Q_m+\Lambda^2 Q_m \times  Q_\epsilon\right] \nn\\
\chi_{\C},\chi_{[\ell_1\ell_2]},\bar{\chi}_{\C}:&& -V\times
V^*\times \left[1+\Lambda^2(Q_\epsilon+Q_m)
+\Lambda^4(Q_\epsilon+Q_m)\right]\nn\\
  w_{\dot{\alpha}},\bar w_{\dot{\alpha}} :&& (V\times W^*+ V^*\times W\times
\Lambda^2 Q_\epsilon )\times (1+\Lambda^2 Q_m) \nn\\
 \mu_{\dot{a}},\bar{\mu}_{\dot{a}}:&& -(V\times W^*+V^*\times W\times \Lambda^2 Q_\epsilon)\times Q_m
\label{fieldtransform} \eeqa with
 \beq Q_\epsilon=Q_1+Q_2 \qquad Q_m=Q_3+Q_4
\eeq
where $Q_\epsilon, Q_m$ are two doublets which are associated to
the contributions to the moduli space of the ${\cal N}=2$ gauge vector
multiplet and
to the matter part respectively \footnote{If we would neglect $B_{3,4}$, $\vec
m$ is the moduli space of ${\cal N}=2$.}.
 The signs account for the right spin statistics. $\Lambda^p$ is
the space of antisymmetrized $p$-vectors such that
$\Lambda^2Q_\epsilon=Q_1Q_2$, $\Lambda^2Q_m=Q_3Q_4$ and $\Lambda^4(Q_\epsilon+Q_m)=Q_1Q_2Q_3Q_4$.

The tangent of the moduli space $\vec{m},\hat{\vec{\chi}}$ is
given by collecting all contributions from (\ref{fieldtransform})
\beqa
&T_{\cal M}
=\left[ V\times V^* (-1+Q_\epsilon-\Lambda^2Q_\epsilon)+ V\times
W^*+V^*\times W\times \Lambda^2 Q_\epsilon\right](1-Q_m+\Lambda^2
Q_m) \nn\\
&=\left[ V\times V^* (-1+Q_\epsilon-\Lambda^2Q_\epsilon)+ V\times
W^*+V^*\times W\times \Lambda^2 Q_\epsilon\right](1-T_m)+ {\rm
h.c.}
\label{msp}
\eeqa
with $T_m=e^{i m}$.
The mathematical origin of (\ref{msp}) is
explained in Appendix C of \cite{Bruzzo:2002xf}. Here we give a more
intuitive argument. All the terms in squared brackets are the
contribution of the ${\cal N}=2$ sector of ${\cal N}=4$. The
ADHM constraints and $U(k)$ invariance are
represented by the negative terms on the r.h.s. of (\ref{msp}) and
they are $2k^2=2({\rm dim}\, V)^2$ complex conditions.\footnote{Moduli spaces are
constructed as quotient spaces. One starts from a flat space to
later impose some constraints and to mod out by a symmetry group.
For the ${\cal N}=2$ case the number of constraints is usually
taken to be $3k^2$ (one real and 1 complex) and the symmetry group
$U(k)$. After the imposition of the constraints the manifold which
is obtained is
not complex anymore since its dimension is odd.
After modding out the $U(k)$ symmetries the
final manifold (the moduli space) is complex. To avoid the real constraint to break
the complex structure in intermediate steps, it can be omitted altogether
at the price of modifying the group of symmetries of the construction to be
$GL(k)$\cite{Hitchin:1986ea}. This is the case here.}. The ambient space
$\C^{2kN+2k^2}$ is given by the positive
terms inside the squared brackets. The total tangent space is thus a
$2kN$-complex dimensional hypersurface embedded in
$\C^{2kN+2k^2}$. Terms proportional to $T_m$ correspond
to the contribution of the matter hypermultiplet in ${\cal N}=2^*$.
They are a copy of the ${\cal N}=2$ terms but with the opposite spin statistics
(as it can be seen already in (\ref{brs})).

The physical content of the localization
formula can be now made precise
by noticing that the result (\ref{locth}) follows by first expanding the action (\ref{N=4}) up to
quadratic order in the fields around a vacuum characterized by $\phi_0$ and
then performing the Gaussian integrations.
To this order we can indeed set ${\cal E}=0$ and the action becomes
quadratic. The vacua are defined by the critical point equations
$$
Q^2 \vec{m}=\delta_{\phi,a,\epsilon} \vec{m}=0
$$ with
$\hat{\vec{H}}=\hat{\vec{\chi}}=\vec{\Psi}=0$.
Notice that after having diagonalized the field $\phi$ and having
performed the Gaussian integration over the remaining fields, we
are left with a complex integral whose poles are given by
the fixed point equations.
Therefore the sum over $\phi_0$ in (\ref{locth}) computes the residue.
In components
\beqa \label{critical}
\delta_{\phi,a,\epsilon} B_{\ell,s s'} &=&(\phi(s)-\phi(s^\prime)
+\epsilon_{\ell})\, B_{\ell,s s'}=0 \qquad \ell=1,2 \nn\\
\delta_{\phi,a,\epsilon} I &=&(\phi(s)-a_\alpha)\, I_{s\alpha}=0 \nn\\
\delta_{\phi,a,\epsilon} J &=& (-\phi(s)+a_\alpha+\epsilon)\, J_{\alpha s}=0
\eeqa
with $B_3=B_4=0$.

The solutions of (\ref{critical}) can be put in one
to one correspondence with a set of $N$ Young tableaux $(Y_1,\ldots Y_N)$ with
$k=\sum_\alpha k_\alpha$ boxes distributed between the $Y_\alpha$'s.
The boxes in a $ Y_\alpha$ diagram are
labelled either by the
instanton index $s$ running over all boxes $s\in Y_\alpha=1,\ldots,k_\alpha$ or by a pair of integers
$i_\alpha,j_\alpha$
denoting the horizontal and vertical
position respectively in the Young diagram.

 For simplicity we will set from now on $\epsilon_1=-\epsilon_2=\hbar$.
This corresponds to consider self-dual gravitational backgrounds.
The explicit solutions to (\ref{critical}) can then be written as
\beqa \label{solcritical}
\phi(s\in Y_\alpha) &=& \phi_{i_\alpha j_\alpha} =
a_{\alpha}+(i_\alpha-1)\epsilon_1+(j_\alpha-1)\epsilon_2=a_{\alpha}+(i_\alpha-j_\alpha)\hbar
\eeqa
and $J=B_\ell=I=0$ except
for the components $B_{1(i_\alpha j_\alpha),(i_\alpha+1 j_\alpha)}$,
$B_{2(i_\alpha j_\alpha),(i_\alpha j_\alpha+1)},\, I_{\alpha,11}$
$\alpha=1,\ldots,N$.
At the critical points the spaces $V, W, Q_\ell$ become $T_{\hbar}=e^{i \hbar}$, $T_m=e^{i m}$
and $T_{a_\alpha}=e^{i a_\alpha}$ modules allowing
the decomposition
\beqa
&& V=\sum_{s\in Y_\alpha} e^{i\phi(s)}= \sum_{s\in Y_\alpha}  T_{a_\alpha} T_{\hbar}^{i_\alpha-j_\alpha}\qquad
W=\sum_{\alpha =1}^{N} T_{a_\alpha}\nn\\
&& Q_1=Q_2^*=T_{\hbar}  \qquad  Q_3=Q_4^*=T_m\label{cinqueapp}
\eeqa

Plugging (\ref{cinqueapp}) in (\ref{msp}) one finds after a long but straightforward algebra
the character
\beq
\chi=(2-T_m-T_m^{-1})\left(\sum_{\alpha,\beta=1}^N \sum_{s\in Y_\alpha}
T_{E_{\alpha\beta}(s)}+{\rm h.c.}\right)
\label{traceh}
\eeq
with $T_{E_{\alpha\beta}(s)}\equiv e^{i E_{\alpha\beta}(s)}$ given in terms of
\beq
 E_{\alpha\beta}(s) =
a_{\alpha\beta}+\hbar\, \ell_{\alpha\beta}(s) \eeq and
$\ell_{\alpha\beta}(s)$ is the "length of the hook stretched
between the tableaux $Y_\alpha$, $Y_\beta$" and centered on the
box $s\in Y_\alpha$. More precisely $\ell_{\alpha\beta}(s)$ is the
standard hook length for a box $s\in Y_\alpha$ (the number of
white circles in Fig.1) plus the vertical difference
$\nu_\beta(s)-\nu_\alpha(s)$ between the top ends of the columns
of $Y_\alpha$, $Y_\beta$ to which $s$ belongs to (the number of
black circles in fig. 1). Notice that when $s\in Y_\alpha$ lies
outside of $Y_\beta$, the hook length can be negative since the
vertical distance between the two tableaux can be negative. In
particular  if $s$ lies outside on the right of
diagram $Y_\beta$ then $\nu_\beta=0$.
See Appendix C of \cite{Bruzzo:2002xf} for a more detailed explanation of the
mathematical meaning of (\ref{traceh}).
%*************************************************************************
\begin{figure}[ht]
\label{figura}
\setlength{\unitlength}{2mm}
\begin{center}
\begin{picture}(50,25)(-5,-20)
% 1 riga
\put(0,0){\dashbox{.2}(3,3)}
%2 riga
\put(0,-3){\dashbox{.2}(3,3)}\put(3,-3){\dashbox{.2}(3,3){$\bullet$}}\put(6,-3){\dashbox{.2}(3,3)}
\put(9,-3){\dashbox{.2}(3,3)}\put(12,-3){\dashbox{.2}(3,3)}
%3 riga
\put(0,-6){\framebox
(3,3)}\put(3,-6){\dashbox{.2}(3,3){$\bullet$}}\put(6,-6){\dashbox{.2}(3,3)}
\put(9,-6){\dashbox{.2}(3,3)}\put(12,-6){\dashbox{.2}(3,3)}
%4 riga
\put(0,-9){\framebox (3,3)}\put(3,-9){\framebox
(3,3){$\circ$}}\put(6,-9){\dashbox{.2}(3,3)}
\put(9,-9){\dashbox{.2}(3,3)}\put(12,-9){\dashbox{.2}(3,3)}\put(15,-9){\dashbox{.2}(3,3)}
%5 riga
\put(0,-12){\framebox (3,3)}\put(3,-12){\framebox
(3,3){$\circ$}}\put(6,-12){\framebox (3,3)}
\put(9,-12){\dashbox{.2}(3,3)}\put(12,-12){\dashbox{.2}(3,3)}\put(15,-12){\dashbox{.2}(3,3)}
%6 riga
\put(0,-15){\framebox (3,3)}\put(3,-15){\framebox
(3,3){$\circ_s$}}\put(6,-15){\framebox (3,3){$\circ$}}
\put(9,-15){\framebox
(3,3){$\circ$}}\put(12,-15){\dashbox{.2}(3,3)}
\put(15,-15){\dashbox{.2}(3,3)}
%7 riga
\put(0,-18){\framebox (3,3)}\put(3,-18){\framebox (3,3)}\put(6,-18){\framebox (3,3)}
\put(9,-18){\framebox (3,3)}\put(12,-18){\dashbox{.2}(3,3)}\put(15,-18){\dashbox{.2}(3,3)}

%
%% 1 riga
%\put(25,0){\framebox(3,3)}
%%2 riga
%\put(25,-3){\framebox(3,3)}\put(28,-3){\framebox(3,3)}\put(31,-3){\framebox(3,3)}
%\put(34,-3){\framebox(3,3)}\put(37,-3){\framebox(3,3)}
%%3 riga
%\put(25,-6){\dashbox{.2}(3,3)}\put(28,-6){\framebox(3,3)}\put(31,-6){\framebox(3,3)}
%\put(34,-6){\framebox(3,3){$0_s$}}\put(37,-6){\framebox(3,3){$\oplus$}}
%%{$\oplus$}}
%%4 riga
%\put(25,-9){\dashbox{.2}(3,3)}\put(28,-9){\dashbox{.2}(3,3)}\put(31,-9){\framebox(3,3) }
%\put(34,-9){\framebox(3,3){$\ominus$}}\put(37,-9){\framebox(3,3) }\put(40,-9)
%{\framebox(3,3) }
%%5 riga
%\put(25,-12){\dashbox{.2}(3,3)}\put(28,-12){\dashbox{.2}(3,3)}\put(31,-12){\dashbox{.2}(3,3) }
%\put(34,-12){\framebox(3,3){$\ominus$}}\put(37,-12){\framebox(3,3)}\put(40,-12){\framebox(3,3)}
%%6 riga
%\put(25,-15){\dashbox{.2}(3,3)}\put(28,-15){\dashbox{.2}(3,3)}\put(31,-15){\dashbox{.2}(3,3) }
%\put(34,-15){\dashbox{.2}(3,3)}\put(37,-15){\framebox(3,3)}
%\put(40,-15){\framebox(3,3)}
%%7 riga
%\put(25,-18){\dashbox{.2}(3,3)}\put(28,-18){\dashbox{.2}(3,3)}\put(31,-18){\dashbox{.2}(3,3) }
%\put(34,-18){\dashbox{.2}(3,3)}\put(37,-18){\framebox(3,3)}\put(40,-18){\framebox(3,3)}
\end{picture}
\caption{Two generic Young diagrams denoted by the indices $\alpha, \beta$ in the main text. In the figures
the diagram $Y_\alpha$ where the box "s" belongs to is always displayed with solid lines. The hook
starts on box "s" a run horizontally till the end of diagram $Y_\alpha$ and vertically till the top
end of $Y_\beta$. The two tableaux are depicted on top of each other
so that in the picture in the left side the solid diagram $Y_\alpha$ contain both solid and dashed boxes.}
\end{center}
\end{figure}
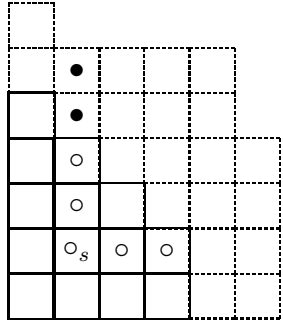
%***************************************************************************

The exponents in (\ref{traceh}) are the eigenvalues of
the operator ${\cal L}_{\phi_0}$
which enter our localization formula (\ref{locth}).
The partition function is then given by replacing the sum $\chi=\sum_j e^{i \lambda_j}$ in (\ref{traceh})
by a product over the eigenvalues $\lambda_j$ leading to
\beqa
Z_{k}&=& \sum_{\phi_0} {1\over {\rm Sdet}^{1/2} {\cal L}_{\phi_0}}=
\sum_{ Y=\{ Y_\alpha \}} \, {\cal Z}_Y\nn\\
{\cal Z}_Y &=&
  \prod_{\alpha,\beta=1}^n
\prod_{s\in
Y_{\alpha}}{(E_{\alpha\beta}(s)+m)(E_{\alpha\beta}(s)-m)\over
E_{\alpha\beta}(s)^2}
\label{inth}
\eeqa
with $k=|Y|$ the instanton number.

It is  convenient to introduce the notation \beqa
f(x)&=&{(x-m)(x+m)\over  x^2} \nn\\
S_\alpha(x)&=&
%\prod_{\beta \neq \alpha} \, f(x-a_{\beta})
=\prod_{\beta \neq \alpha}\, \left[{(x-a_{\beta} +m)(x-a_{\beta} -m)\over
(x-a_{\beta})^2}\right] \label{fst} \eeqa
 In terms of these functions the contributions coming from the tableaux
with $k=1,2$ can be written as \cite{Bruzzo:2002xf}
\beqa
{\cal Z}_{\tinyyoung{\hfil}}&=&q\,\sum_\alpha f(\hbar) S_\alpha(a_\alpha)\nn\\
{\cal Z}_{\tinyyoung{\hfil},\tinyyoung{\hfil}}&=& \ft12 \,q^2\,
\sum_{\alpha\neq \beta}\, S_\alpha(a_\alpha) S_\beta(a_\beta) \,
f(a_{\alpha\beta}+\hbar)f(a_{\alpha\beta}-\hbar){f(\hbar)^2 \over f(a_{\alpha\beta})^2}\nn\\
{\cal Z}_{\tinyyoung{\hfil}\tinyyoung{\hfil}}&=&  q^2\,\sum_\alpha
f(\hbar) f(2\hbar) S_\alpha(a_\alpha) S_\alpha(a_\alpha+h)
\label{zr4}
\eeqa
 with $S_\alpha=S_\alpha(a_\alpha)$ and ${\cal Z}_{\tinyyoung{\hfil,\hfil}}$ given by
${\cal Z}_{\tinyyoung{\hfil}\tinyyoung{\hfil}}$
 with $\hbar\to -\hbar$. In a similar way the $k$-instanton partition function
  can be written in terms
 of a products of $k$ $S_\alpha(x)$'s. %These characteristic functions

The multi instanton partition function  $ Z(q)=\sum_k Z_k q^k$
determines the prepotential ${\cal F}(q)=\sum_k{\cal F}_kq^k$  via the relation
\beq
{\cal F}(q)\equiv \lim_{\hbar\to 0}\,{\cal F}(q,\hbar)=
\lim_{\hbar\to 0}\,\hbar^2 ln \, Z(q)
\label{prep}
\eeq
The
general function ${\cal F}(q,\hbar)$ encodes the gravitational
corrections to the ${\cal N}=2^*$ superpotential. %We remind the reader that for $SU$ and $SO$ gauge groups
%the relation ${\cal F}_1=\sum_\alpha S_\alpha(a_\alpha)$ holds, while for the gauge group $Sp$ we have
%${\cal F}_1=-2[\bar S_0(0)]^{1\over 2}$ with
%\beq
%S(x)={\bar S_0(x)\over x^4}
%\label{testfunc1}
%\eeq
\section{Quiver gauge theories}
\setcounter{equation}{0}
\subsection{Generalities}
Quiver gauge theories are defined by projecting ${\cal N}=4$ SYM
with the orbifold group $\Gamma$ now embedded in the ${\cal
R}$-symmetry group. This group has been already used to introduce
mass deformations parameterized by a $U(1)_m$ (whose action is given by $T_m$)
breaking the original
$SO(6)$ ${\cal R}$-symmetry group of ${\cal N}=4$ down to that of
${\cal N}=2^*$  \cite{Hollowood:1999bm,Fucito:2001ha}. Under the
action of $\Gamma=\Z_p$ the spaces $V, W$ decompose as
\beq
V=\sum_{q=0}^{p-1}\, V_q \qquad
W=\sum_{q=0}^{p-1}\, W_q
\label{decomp}
\eeq with $q$ labelling the
$q^{th}$ irreducible representation $R_q$ of $\Z_p$ under which
the corresponding $k_q={\rm dim} V_q$ D(-1) instantons and
$N_q={\rm dim} W_q$ D3-branes, transform.
The symmetry group $U(k)\times U(N)$ becomes $\prod_{q=0}^{p-1}\,
U(k_q)\times U(N_q)$.

 The orbifold group
generator in $\Gamma=Z_p$ is taken along the $U(1)_m$ \beqa
\Z_p\quad : m \to m+{2\pi\over p}\quad\quad a_\alpha\to
a_\alpha+{2\pi q_\alpha \over  p}
 \label{orbprojm} \eeqa $q_\alpha,\quad \alpha=1,\ldots,N$ can
take integer values between $0$ and $p-1$, specifying the
representations under which the $\alpha^{\rm th}$ D3-brane
transforms.
In particular the integers $N_q$
characterizing the unbroken gauge group $\prod_q U(N_q)$ are given
by the number of times that $q$ appears in $(q_1,q_2,\ldots q_N)$.
The integers $k_q={\rm dim} V_q$ are given by the total number of boxes
in the Young tableaux with $q_\alpha=q$.

 Equivalently in terms of the decompositions (\ref{decomp}) the
 orbifold group action can be written as
\beq
\Z_p\quad : m \to m+{2\pi\over p}\quad\quad V_q\to e^{2i\pi q
\over  p} \, V_q \qquad  W_q\to e^{2i\pi q \over  p} \, W_q
 \label{orbprojm2}
\eeq
The $\Gamma$ invariant component of the tangent space (\ref{msp})
under (\ref{orbprojm2}) can then be written as
\beqa
\label{traceG2}
T_\Gamma&=&\sum_{q=0}^{p-1} \left[ V_q^*\,
V_q(Q_\epsilon-1-\Lambda^2 Q_\epsilon) + W_q^*\,V_q +V_q^*\, W_q\, \, \Lambda^2 Q_\epsilon
\right] \label{vqs}\\&& - T_m\sum_{q=0}^{p-1} \left[ V_q^*\,
V_{q-1}(Q_\epsilon-1-\Lambda^2 Q_\epsilon) + W_q^*\,V_{q-1} +V_q^*\, W_{q-1}\, \,
\Lambda^2 Q_\epsilon \right]+{\rm h.c.} \nn
\eeqa
In particular for
$V=V_0,V_{q\geq 1}=0$, that is when all the instantons sit in
the $q=0$ D3-branes, the $\Gamma$-invariant moduli space (\ref{vqs})
reduces to
\beqa
T_\Gamma&=& \left[ V_0^*\, V_0+ V_0^*\, V_0-
V_0^*\, V_0\,\Lambda^2 Q_\epsilon-V_0^*\, V_0 + W_0^*\, V_0 +V_0^*\, W_0\,
\Lambda^2 Q_\epsilon \right] \nn\\&& - Q_m\left[
   W_{N-1}^*\, V_0 +V_0^*\, W_1\, \,
\Lambda^2 Q_\epsilon \right]\nn
\eeqa
which coincides with the tangent to
the moduli space of an ${\cal N}=2$ SYM theory with gauge group
$U(N_0)$, $N_{1}$ fundamentals and
$N_{p-1}$ anti-fundamentals
(see (3.21) in \cite{Bruzzo:2002xf}). In particular for $p=2$ and
$N_0=N_1$ we find the ${\cal N}=2$ $U(N_0)$ superconformal theory
with $2N_0$ fundamental matter \cite{Bruzzo:2002xf}. The general
case described in (\ref{traceG2}) corresponds to ${\cal N}=2$
gauge theories with product gauge groups $\prod_{q} U(n_q)$ and
bifundamental matter. The character of the $\Gamma$-invariant
tangent space follows from the $\Gamma$-invariant component of
(\ref{traceh}) under (\ref{orbprojm}) and can be written as
\beqa
\chi_{m,\Gamma} &= &\sum_{\alpha,\beta}^N \sum_{s\in Y_\alpha}
T_{E_{\alpha\beta}(s)}\, \left(  2
\,\delta_{q_\alpha,q_\beta}-T_m\,\delta_{q_\alpha,q_{\beta+1}}
-T_m^{-1}\,\delta_{q_\alpha,q_{\beta-1}}\right)+{\rm h.c.}
\label{tracef2}
\eeqa
with $q=0,\ldots p-1$ running over the various factors in the
product gauge group $\prod_q U(n_q)$.

The partition function reads
\beq
 Z_k  =
\sum_{ \{ Y_\alpha \}} \, \prod_{\alpha_q,\beta_q}
\prod_{s\in
Y_{\alpha}}{(E_{\alpha_q \beta_{q+1}}(s)+m)(E_{\alpha_q \beta_{q-1}}(s)-m)\over
E_{\alpha_q \beta_q }(s)^2}
\label{inth2}
\eeq

Notice in particular that
the contribution of the matter fields ( $m$-dependent eigenvalues) turns out to
always be off-diagonal in agreement with the fact that they
always transform in bifundamental representations.

\subsection{The quiver prepotential}

Finally we derive the multi instanton partition functions and
prepotentials describing the low energy physics on quiver gauge
theories descending from ${\cal N}=2^*$ gauge theories via $\Z_p$
projections \footnote{In this chapter we will give the lowest
terms in the expansion of the prepotential using an analytical
method for pedagogical purposes. To check highest order terms,
we use a Mathematica
code.}.

For concreteness we restrict
ourselves to the case $U(N_0)\times U(N_1)$, i.e. $N_q=0$ for
$q>1$. More precisely we consider a ${\cal N}=2$ quiver theory with gauge group
$U(N_0)\times U(N_1)$ and bifundamental hypermultiplets with mass
$m$ in representations $(N_0,\bar{N}_1)$ and $(N_{p-1},\bar{N}_0)$ of
the gauge group.
Notice that $R_1=\bar R_1$ is real for $\Z_2$ but complex for $\Z_p$ with $p>2$.
This implies in particular that matter  in the fundamental and antifundamental
will transform in the same way under $\Z_p$ for $p=2$ and therefore
both fundamental and anti-fundamental matter will survive the $\Z_2$ projection.
On the other hand for $\Z_p$ projections with $p\geq 2$
only the fundamental matter survives the orbifold projection.

We start by consider the $\Z_2$ case.
 It is convenient to split the gauge index $\alpha$ into
$\alpha=r, \hat r$ with $r=1,\ldots N_0$ and
$\hat r=1,\ldots N_1$. This corresponds to
choose two sets of D3-branes transforming in the $R_0$ and $R_1$
representation of $\Z_2$.

We now need to project (\ref{fst}) onto its invariant components.
The projection is given by (\ref{orbprojm}) with $q_r=0, q_{\hat
r}=1$. Given the gauge group $U(N_0)\times U(N_1)$ we see that
under (\ref{orbprojm}) the combinations $a_{r_1 r_2}, a_{\hat{r}_1,\hat{r}_2},
a_{r_1,\hat r_2}\pm m$ are invariant.
 The invariant components in (\ref{fst}) are then given by
 \beqa
 && f(a_{r_1 \hat{r_2}}+x )= (a_{r_1 \hat{r_2}}+x )^2-m^2 \qquad f(x)={1\over x^2} \nn\\
&& f(a_{r_1 r_2}+x )= {1\over (a_{r_1 r_2}+x )^2} \qquad
f(a_{\hat{r_1} \hat{r_2}}+x )= {1\over (a_{\hat{r_1} \hat{r_2}}+x )^2}\nn\\
&& S_{r_1}(x)= {\prod_{\hat{r_2} }\,
 (x-a_{\hat{r_2}})^2-m^2\over
 \prod_{r_2 \neq r_1 }\, (x-a_{r_2})^{2}}
 \qquad S_{\hat{r_1}}(x)= {\prod_{r_2}\,
 (x-a_{r_2} )^2-m^2\over
 \prod_{\hat{r_2} \neq \hat{r_1} }\, (x-a_{ \hat{r_2}})^{2}}
 \label{replace}
\eeqa In (\ref{replace}) the variable $x$ does not transform under
(\ref{orbprojm}). Once again one can verify the formulae
we obtain for $S_{r_1}(x)$,
$S_{\hat{r}_1}(x)$ against those coming from the Seiberg and Witten
curves \cite{Gomez-Reino:2003cp}. As expected, the contributions coming from the
massless ${\cal N}=2$ vector
 multiplet appear in the denominators of (\ref{replace}) and are in the adjoint of
$U(N_0)\times U(N_1)$ (diagonal terms).
The contributions from the massive matter are in the numerators
and appear off-diagonally as bifundamentals.
This can also be seen from the character (\ref{vqs}) where the
term $-T_m$, representing the contribution of the massive
hypermultiplet, is the only term transforming under the orbifold
projection (\ref{orbprojm}). To compensate this action the
corresponding  bilinear in $V_p,W_q$ should transform accordingly.

The partition function is given again by (\ref{zr4})  but now with
$\alpha$ running over $r,\hat{r}$ and $f(x),S_\alpha(x)$ given by
(\ref{replace}).
The prepotential is given by (\ref{prep}) leading to
\beqa
&&{\cal F}=q \,\sum_r S_r+\hat{q}
\,\sum_{\hat r} S_{\hat r}
 +q^2\,\left(\ft14\sum_r \, S_r S^{''}_r+
 \sum_{r_1\neq r_2}\, { 1\over a_{r_1 r_2}^2}\,     S_{r_1} S_{r_2}\right)\nn\\
&&+\hat{q}^2\,\left( \ft14\sum_{\hat r} \, S_{\hat r}
S^{''}_{\hat r} +\sum_{\hat{r_1}\neq \hat{r}_2}\, {1 \over
a_{\hat{r}_1 \hat{r}_2}^2}\, S_{\hat{r}_1}S_{\hat{r}_2}\right)
-2q \hat{q}\sum_{r_1,\hat{r_2}}\, { (a_{r_1
\hat{r}_2}+\hbar)^2+m^2\over (a_{r_1 \hat{r_2}}^2-m^2)^2}\,
S_{r}S_{\hat s}+\ldots
\label{f4m}
\eeqa
where $S_r=S_r(a_r)$.
Here we weight the
instanton contributions by $q=e^{2\pi i k \tau}$, $\hat q=e^{2\pi i
\hat{k} \hat{\tau}}$ with  $\tau,\hat{\tau}$ the coupling
constants of the two gauge groups.

In appendix \ref{aquiv} we present the results for $Z_p$
quivers with $p>1$ corresponding to quiver gauge theories
with a single hypermultiplet in the bifundamental.

 Finally one can consider quivers of $U(N)$ rather
than $SU(N)$ groups.
 As an illustration consider the $\Z_2$-quiver of ${\cal N}=2^*$ SYM with
starting gauge group $U(2)$. There are two possibilities according
to whether we choose $q_1=q_2$ or $q_1\neq q_2$ in (\ref{orbprojm}).
They lead to
quiver gauge theories with gauge groups $U(2)$ and $U(1)^2$
respectively. In the former case the projection (\ref{orbprojm})
remove precisely the massive fields leading to pure ${\cal N}=2$
SYM. In the latter case (with $q_1=0, q_2=1$) one finds
\beqa
Z(q,\hbar)&=&
1+\left({q\over\hbar^2}\right)(8a^2-2m^2)+\left({q\over\hbar^2}\right)^2
(32a^4+2m^4-3m^2\hbar^2+\hbar^4-16a^2m^2-12a^2\hbar^2)+\nn\\
&&\frac{2}{3}\left({q\over\hbar^2}\right)^3
(4a^2-m^2)(32a^4+2m^4-9m^2\hbar^2+7\hbar^4-16a^2m^2-12\hbar^2a^2)+\nn\\
&&\frac{1}{6}\left({q\over\hbar^2}\right)^4
(1024a^8+4m^8-36m^6\hbar^2+71m^4\hbar^4-39m^2\hbar^6+12\hbar^8-1024a^6m^2\nn\\
&&-768a^6\hbar^2
+384a^4m^4-392a^4m^2\hbar^2+752a^4\hbar^4-64a^2m^6+240a^2m^4\hbar^2\nn\\
&&-280a^2m^2\hbar^4-84a^2\hbar^6)+\ldots \label{zquiv}\eeqa and
 \beqa {\cal
F}(q)&=&\left[\left(-8a^2+2m^2\right)q+\left(4a^2+3m^2\right)
q^2+\frac{8}{3}\left(-4 a^2+m^2\right)q^3+\right.\nn\\
&& \left. \left(10a^2+{7 m^2\over 2}\right)q^4+
\frac{12}{5}\left(-4a^2+m^2\right)q^5+
\frac{4}{3}\left(4a^2+3m^2\right)q^6 +O(q^7)\right]\nn\\
&&-\hbar^2 \left[q^2+\frac{3}{2}q^4+\frac{4}{3}q^6+ O(q^7)
\right]
\eeqa
Remarkably the expansion in $\hbar$ stops now at
$\hbar^2$  i.e. instanton corrections to gravitational terms
${\cal F}_g \hbar^{2g}$ vanish beyond one loop $g\geq 1$. This is
clearly peculiar for this simple situation where only powers of
$\hbar$ are left in the denominator in (\ref{zquiv}) due to the
orbifold projection. This implies in particular that ${\cal
F}_k(a,m,\hbar)$ is given by a homogenous polynomial of
$a,m,\hbar$ of order two.

\section{Orientifolds}
\setcounter{equation}{0}
 SYM theories with gauge group SO/Sp can be studied by including
an O3 orientifold plane in the D3-D(-1) system. The orientifold plane can be thought of as a
 mirror where the open strings reflect, losing their
 orientation.
%***************************************************************************
\begin{figure}
\begin{center}
\begin{pspicture}(0,0)(15,7)
% 2^ brana
\psline[linestyle=dashed](2.5,1)(3,.75)
\psline(3,.75)(4.5,0)
\psline[linestyle=dashed](2.5,1)(2.5,4.25)
\psline(2.5,4.25)(2.5,5)
\psline(4.5,0)(4.5,4)
\psline(2.5,5)(4.5,4)
% 1^ brana
\pspolygon(1,1)(3,0)(3,4)(1,5)
%piano o3
\pspolygon[origin={-6,0}](1,1)(3,0)(3,4)(1,5)
% 1^ brana mirror
\pspolygon[origin={-10,0}](1,1)(3,0)(3,4)(1,5)
%2^ brana mirror
\psline[linestyle=dashed](12.5,1)(13,.75)
\psline(13,.75)(14.5,0)
\psline[linestyle=dashed](12.5,1)(12.5,4.25)
\psline(12.5,4.25)(12.5,5)
\psline(14.5,0)(14.5,4)
\psline(12.5,5)(14.5,4)
% stringa aperta(5,3)
\pscurve(3.5,2)(6.5,1.5)(8,2)(10,2)(11.8,1.5)
\rput(2,-.5){$a_2$}
\rput(4.0,-.5){$a_1$}
\rput(8,5.5){O3-Plane}
\rput(.5,5.5){D3-brane}\rput(3.5,5.5){D3-Brane}
\rput(12,-.5){$-a_1$}
\rput(14.0,-.5){$-a_2$}
\psbezier{->}(.2,1.2)(.8,2.3)(1,1.5)(1.3,2)
\rput(-.5,1){D(-1)-Brane}
%d-1 sulle prime brane
\psdots(3.5,1.5)
\psdots(3.5,2)
\psdots(3.5,2.5)
\psdots(4,1.5)
\psdots(1.5,1.5)
\psdots(1.5,2)
%d-1 sulle brane mirror
\psdots(11.3,1.5)
\psdots(11.3,2)
\psdots(11.8,1.5)
\psdots(12.3,1.5)
\psdots(13.5,1.5)
\psdots(14,1.5)
\end{pspicture}
\vskip.5cm
\caption{D(-1)-D3 branes in the presence of an orientifold plane.
We have also drawn a D(-1)-D(-1) open string. See the text
for a more detailed explanation of the figure.}
\end{center}
\end{figure}
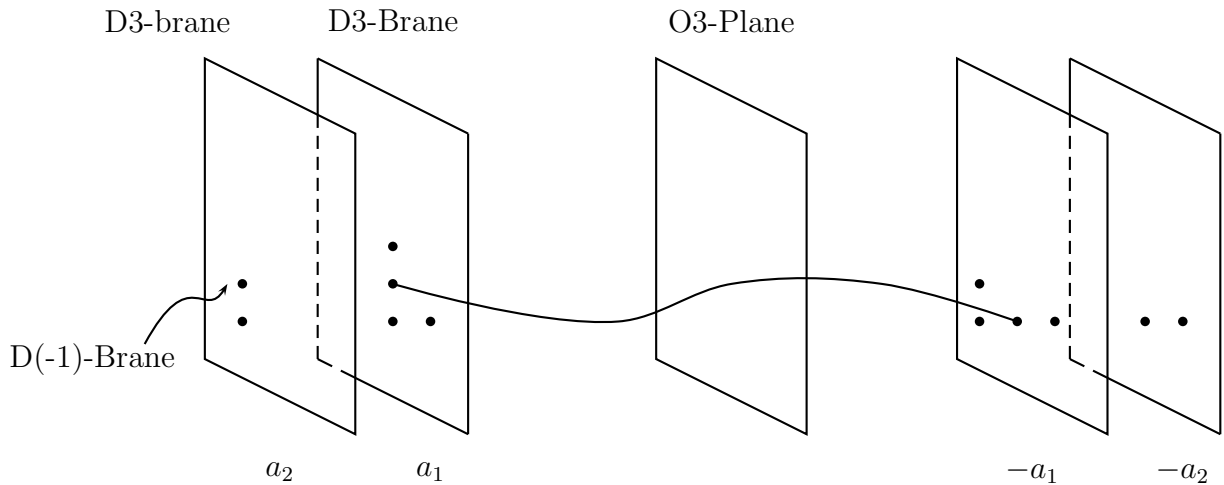
%**************************************************
Like in the orbifold case the character of the tangent space to
the multi-instanton moduli manifold can be found via a projection.
The projection operator is $(1\mp\Omega \Z_2)/2$. Besides a
reflection of the six transverse coordinates to the O3 plane, we
have a conjugation (worldsheet parity) on the Chan-Paton indices.
In terms of the previously introduced $V, W$ and $Q_\epsilon$ spaces, we have
\beq
\Omega \Z_2:  V \to  V^*  \qquad  W_\alpha \to W_\alpha^*
\qquad Q_\epsilon\to -Q_\epsilon
\eeq
The choices $\mp\Omega \Z_2$ correspond to  $SO/Sp$ gauge theories.
The gauge symmetries of the D(-1)-D3-O3 system become $SO(N)\times Sp(2k)$
or $Sp(2n)\times SO(K)$.
After the v.e.v.'s are
turned on, the D3-branes distribute over the $\phi$-plane in
mirror-image pairs $a_{\alpha}=a_1,\ldots a_n,-a_1,\ldots,-a_n$ for
$SO(2n)/Sp(2n)$ while for $SO(2n+1)$ an additional D3-brane sits at the
origin (($a_{2n+1}=0$) on top of the orientifold plane. In a similar way instantons
distribute symmetrically between the two sides of the O3 mirror
\footnote{
No D(-1)-branes are allowed to be superposed to the $(2n+1)^{th}$ D3-brane in the $SO(2n+1)$ case.}.
This corresponds to distributing instantons over a particular set
of Young tableaux of the type
\beqa
Y_{SO(2n)}=(Y_1,\ldots Y_n,Y_1^T,\ldots Y_n^T)\nn\\
Y_{SO(2n+1)}=(Y_1,\ldots Y_n,Y_1^T,\ldots Y_n^T, \emptyset)\nn\\
Y_{Sp(2n)}=(Y_1,\ldots Y_n,Y_1^T,\ldots Y_n^T, Y_\nu)
\label{mir}
\eeqa
As it is stressed in Fig.2, the second half of the Young tableaux are conjugate to the first half i.e.
the number of boxes in rows and columns are interchanged.
The $Sp(2n)$ gauge theory admits special solutions to (\ref{critical}) which are not present in
the parent $SU(2n)$ gauge theory: we call these solutions "fractional instantons".
They are distributed symmetrically around the O3-plane at the origin and outside of any D3-brane.
Also these solutions can be put in relation with particular
types of tableaux $Y_\nu$ (see \cite{Marino:2004cn} for details).
We will start by considering the case $Y_\nu=\emptyset$ and include the contributions
of fractional instantons only at the end.

The choice (\ref{mir}) implies that the spaces $V$ and $W$
are $2k$ and $2n$ dimensional respectively and real  and therefore
 the projection under $(1\mp \Omega\Z_2)/2$ makes sense. More precisely
\beqa V&=&\sum_{\alpha=1}^{2n} \sum_{s\in Y_\alpha} e^{i\phi(s)}= \sum_{r=1}^{n}\sum_{s\in Y_r}
e^{i\phi(s)}
+\sum_{r=1}^{n}\sum_{s\in Y_{\hat r}} e^{i\phi(s)}=V_\phi + V_\phi^*\nn\\
W&=&\sum_{\alpha =1}^{2n} T_{a_\alpha}=\sum_{r =1}^n e^{i
a_r}+\sum_{\hat{r} =1}^n e^{ia_{\hat{r}}}=W_+ + W_+^*
\label{spaces} \eeqa

Non-diagonal terms like $V\times W$ or  $V_{\phi(s)}\times
V_{\phi(s')}$ with $s \neq s'$ come always in $\pm$ pairs
and therefore only one half of them survive the projection. The
$(1\mp\Omega \Z_2)/2$ invariant tangent space can then be written
by applying the projection operator to (\ref{msp}) (in which the
$V, W$ spaces are those in (\ref{spaces})) and taking the trace.
The term $\ft12 {\rm Tr} \,1$ (the annulus)
gives half the contribution of $SU(2n)$
evaluated on the symmetric configurations (\ref{mir}) and can be computed
using (\ref{traceh}).
On the other hand
the trace $\ft12 {\rm Tr}\Omega \Z_2$ (Moebius
strip)\footnote{To stress the similarity with the construction
of the open string partition function,
in parenthesis we have given the names of the analogous terms which
arise in that case.} receives
only contributions from diagonal terms since the contributions of
$\pm$ paired states cancel against each other.
 Collecting the two contribtuions for the pure ${\cal N}=2$ SYM case one finds
\beqa
T_{\cal
M}=\left[ \ft12 (V^2 \mp V_{2\phi})\times Q_\epsilon-\ft12 (V^2
\pm V_{2\phi})\times (1+\Lambda^2 Q_\epsilon) +\ft12 V\times W
\times(1+\Lambda^2 Q_\epsilon)\right]+{\rm h.c.}
\label{tangentspso}
\eeqa
with upper/lowers signs corresponding from now on to
$SO(N)$/$Sp(N)$ gauge groups respectively and
\beqa
V_{2\phi}&=&\sum_{\alpha=1}^{2n}\sum_{s\in
Y_\alpha} e^{2i\phi(s)}= \sum_{r=1}^{n}\sum_{s\in Y_r} \left(e^{2i\phi(s)}+e^{-2i\phi(s)}\right)
%+\sum_{r=1}^{n}\sum_{s\in Y_{\hat r}} e^{2i\phi(s)}
%=V_{2\phi,+} + V_{2\phi,+}^*
\nn\\
\phi ({s\in Y_\alpha})  &=&
 a_{\alpha}+(i_\alpha-j_\alpha)\hbar
\label{tmpure}
\eeqa
 For instance, specifying to $SO(N)$,
 the ADHM constraints (the terms multiplied by
$\Lambda^2 Q_\epsilon$) correspond to matrices
in the symmetric $ k(2k+1)$ representation (the adjoint of
$Sp(2k)$) while the $B$-moduli (which are the terms multiplied by
$Q_\epsilon$) come in the antisymmetric representation of $Sp(2k)$ as
expected. This is in agreement with the standard ADHM construction
for $SO(N)$ groups. The $Sp(N)$ case
behaves analogously with the symmetric and
antisymmetric representations exchanged.

The D-instanton character is given by
\beqa
\chi &=&
\,\sum_{\alpha=1}^{2n} \sum_{s\in Y_\alpha} \left[ \sum_{\beta}^{2n}
T_{E_{\alpha \beta}(s)}\mp \ft12
\,e^{2i\phi(s)}(2+T_\hbar+T_\hbar^{-1})\right]
 +{\rm h.c.}
\label{charor}
\eeqa
  The sum over $\alpha,\beta$ splits into four terms according to whether
 we choose $\alpha,\beta$ between the set $r_{1,2}=1,\ldots n\equiv \ft12 N$ or their
 images $\hat{r}_{1,2}$. In particular notice that while $E_{r_1 r_2}(s)=-E_{\hat{r_1} \hat{r_2}}$ is
a function of the difference $a_{r_1 r_2}\equiv a_{r_1}- a_{r_2}$,
the contributions $E_{r_1 \hat{r_2}}(s)$,$E_{r_1 \hat{r_2}}(s)$
coming from mixed pair of tableaux are functions of the sum
$a_{r_1 \hat{r_2}}\equiv a_{r_1}+ a_{r_2}$.
 The result can then be written as
 \beqa
\chi &=&
\sum_{r_1=1}^n \sum_{s \in Y_{r_1}}\left[ \sum_{r_2=1}^n
\left(2\,T_{E_{r_1 r_2}(s)}+ T_{E_{r_1 \hat{r_2}}} +
T_{E_{\hat{r_1} r_2}}\right)
\mp \,e^{2i\phi(s)}(2+T_\hbar+T_\hbar^{-1}) \right]+{\rm h.c.}
\label{trorb}
\eeqa
Here and in the following we write
$\sum_{s \in Y_{r_1}} \,T_{E_{\hat{r_1} r_2}}$ to indicate
$\sum_{s \in Y_{\hat{r}_1}} \,T_{E_{\hat{r_1} r_2}}$. The diagram
$Y_\alpha$ to which the box "$s$" belongs will be unambiguously specified
by the first label of $E_{\alpha \beta}$.
Specifying to the $SO(2n)$ case, one then finds
\beqa
{\cal Z}^{SO(2n)}_{2k} &=&
  \prod_{r_1=1}^n
\prod_{s\in Y_{r_1}}\, \left[ 4\,\phi(s)^2
\,(4\,\phi(s)^2-\hbar^2)\, \prod_{r_2=1}^n {1\over E_{r_1
r_2}(s)^2 \,E_{r_1 \hat{r}_2}(s)\, E_{\hat{r}_1
r_2}(s)}\right]\label{res}
\eeqa
For $SO(2n+1)$ we have an
additional contribution from the strings
ending on the brane at the origin, i.e. $\prod_{s\in Y_{r_1}} |E_{r_1,n+1} E_{\hat{r}_1,n+1}| =
\prod_{s\in Y_{r_1}}  \phi(s)^2$, so
 that one finds
\beq
{\cal Z}^{SO(2n+1)}_{2k} ={\cal Z}^{SO(2n)}_{2k}\,
  \prod_{r_1=1}^n
\prod_{s\in Y_{r_1}}{1 \over \phi(s)^2}
\label{s02n1}
\eeq

To check these results (and those we will find later), we find it
convenient to compute the lowest instanton contribution.
Specifying to a single instanton pair ($k=1$)   one finds
\beqa
 Z_2^{SO(2n)} &=& {1\over \hbar^2}\,\sum_{r_1} 4 a_{r_1}^2\,
  \prod_{r_2\neq r_1} {1 \over (a_{r_1}^2-a_{r_2}^2)^2}\nn\\
Z_2^{SO(2n+1)} &=& {4\over \hbar^2}\,\sum_{r_1}
  \prod_{r_2\neq r_1} {1 \over (a_{r_1}^2-a_{r_2}^2)^2}
\label{sfunc} 
\eeqa
 (\ref{sfunc}) can be rewritten as
$$
Z_2^{SO(N)}={1\over \hbar^2} \,\sum_{r_1=1}^n  S_{r_1}^{SO(N)}
(a_{r_1})
$$
 with \beqa S_{r_1}^{SO(2n)}(x) &=& {(2x)^4\over
(x+a_{r_1})^2}
  \prod_{r_2\neq r_1} {1 \over (x^2-a_{r_2}^2)^2}\nn\\
S_{r_1}^{SO(2n+1)}(x) &=& {(2x)^4\over x^2(x+a_{r_1})^2}
  \prod_{r_2\neq r_1} {1 \over (x^2-a_{r_2}^2)^2}
\label{sfunc11} \eeqa
In a similar way, ${\cal Z}_{2k}$ contributions can be written as products
of $k$ functions $S_{r_1}^{SO(2n)}(x)$ (see \cite{Marino:2004cn} for details).
 The appearance of functions of the type of (\ref{sfunc}) is typical in the 
 analysis \'a la Seiberg-Witten.
They are related to the polynomial $S(x)$ entering in the
Seiberg-Witten curve via
\beq S^G(x)={S^G_r(x)\over (x-a_r)^2} \label{testfunc}
\eeq
leading to \footnote{For $SU(n)$ one finds
 $S^{SU(n)}(x) =
  \prod_{r=1}^n {1 \over (x-a_{r})^2}$}
\beqa
S^{SO(2n)}(x) &=& (2x)^4\,
  \prod_{r=1}^n {1 \over (x^2-a_{r}^2)^2}\nn\\
S^{SO(2n+1)}(x) &=& {(2x)^4\over x^2}
  \prod_{r=1}^n {1 \over (x^2-a_{r}^2)^2}
\label{sfunc2} \eeqa
This can be compared with the results of Table 3 of \cite{Ennes:1999fb}
\footnote{With respect to Table 3 of \cite{Ennes:1999fb} our results is globally multiplied
by $2^{4k}$.} (see also \cite{D'Hoker:1996mu}).
 Remarkably the first few instanton contributions
contains already enough information to
reconstruct from it the full Seiberg-Witten curve. This will be
used later on to test our results in the case of massive deformations.

The contributions of regular instantons to $Sp(2n)$ can be found from the
$SO(2n)$ case by flipping the sign of the $V_{2\phi}$ term in (\ref{tangentspso})
leading to
 \beqa
{\cal Z}^{Sp(2n)}_{2k,0} &=&
  \prod_{r_1=1}^n
\prod_{s\in Y_{r_1}}\, \left[ {1\over 4\,\phi(s)^2
\,(4\,\phi(s)^2-\hbar^2)}\, \prod_{r_2=1}^n {1\over E_{r_1
r_2}(s)^2 \,E_{r_1 \hat{r}_2}(s)\, E_{\hat{r}_1
r_2}(s)}\right]
\label{spn} 
\eeqa 
The second subscript in ${\cal
Z}^{Sp(2n)}_{2k,0}$ reminds the reader that only instanton
configurations with no fractional instantons ($Y_\nu=\emptyset$) are taken
into account.
In general we will write
\beq
Z^{Sp(2n)}(q)=\sum_{K} {\cal Z}^{Sp(2n)}_{K}=\sum_{k,m}\, {\cal Z}^{Sp(2n)}_{2k,m} q^{2k+m}
\eeq
with $m$ counting the number of fractional instantons.
Fractional instantons can be included by replacing
$V\to V+V_\nu$ in (\ref{tmpure}), with $V_\nu$ describing the
fractional instanton configuration.
  The extra contributions
to the character (\ref{tmpure}) are given by \beqa
\delta T_{\cal M}&=&\ft12 V_{2\nu}\times(Q+1+\Lambda^2 Q)+\ft12(V_\nu^2+2 V V_\nu)\times(Q-1-\Lambda^2 Q)\nn\\
&&+\ft12 V_\nu\times W\times (1+\Lambda^2 Q) \eeqa A single
fractional instanton is constrained to stay at the origin, two
must be symmetrically arranged so to be centered at $(x,y)=(\pm
\ft{\hbar}{2},0)$ or $(x,y)=(0,\pm
\ft{\hbar}{2})$, etc. We denote the corresponding $V_\nu$ and $V_{2\nu}$
by $V_\bullet$, $V_{\bullet\bullet}$ and $V_{2\bullet}$,
$V_{2\bullet\bullet}$ respectively. Explicitly
\beqa
V_\bullet &=& 1 \qquad~~~~~~~~~~  V_{2\bullet}=1 \nn\\
V_{\bullet\bullet} &=& T_{\hbar\over 2}+T_{-{\hbar\over 2}}
\qquad
V_{2\bullet\bullet} = T_{\hbar}+T_{-{\hbar}} \, .\nn \eeqa
Plugging in the character one finds
\beqa
\delta \chi_\bullet &=& 2\, T_{\hbar}+(2\, T_{\hbar}-2)\,V+W+{\rm h.c.}\nn\\
 \delta \chi_{\bullet\bullet} &=& 2\, T_{\hbar}+2 \,T_{2\hbar}+ 2\,T_{\hbar\over 2}\,(T_{\hbar}-T_{-{\hbar}})\, V
  +2\, T_{\hbar\over 2}\, W  +{\rm h.c.}
\eeqa
and
\beqa
  {\cal Z}^{Sp(2n)}_{2k,1}
&=&  \ft12\, {\cal Z}^{Sp(2n)}_{2k,0} {1\over
\hbar^2}\,\prod_{r_1=1}^n\left[ {1\over (-a_{r_1}^2)} \prod_{s\in
Y_{r_1}} {\phi^4(s) \over(\phi^2(s)-\hbar^2)^2} \right]\nn\\
{\cal Z}^{Sp(2n)}_{2k,2} &=& \ft{2}{2^2}\,  {\cal Z}^{Sp(2n)}_{2k,0} {1\over 4
\hbar^4}\,\prod_{r_1=1}^n\left[ {1\over (a_{r_1}^2-{\hbar^2\over
4})^2} \prod_{s\in Y_{r_1}} {(\phi^2(s)-{\hbar^2\over
4})^2\over (\phi^2(s)-{9\hbar^2\over 4})^2} \right]\label{zsp}
\eeqa
 In (\ref{zsp}) we weight the contribution of $m$-fractional instantons by $2^{-m}$.
The extra factor of two for $m=2$ takes into account the identical contributions
coming from the two choices of $Y_\nu$, i.e. $(x,y)=(\pm {\hbar\over 2},0)$ or $(x,y)=(0,\pm {\hbar\over 2})$.
For $k=1,2$ one finds
 \beqa
 Z^{Sp(2n)}_{0,1}
&=& {1\over 2\hbar^2}\,\prod_{r_1=1}^n  {1\over
(-a_{r_1}^2)}\label{funcs2}\nn\\
Z^{Sp(2n)}_{2,0}&=& \sum_r {1\over \hbar^2} {1\over 4 a_{r_1}^2(4 a_{r_1}^2-\hbar^2)^2}
  \prod_{r_2\neq r_1} {1 \over (a_{r_1}^2-a_{r_2}^2)^2}\nn\\
  Z^{Sp(2n)}_{0,2}
&=& {1\over
8 \hbar^4}\,\prod_{r_1=1}^n{1\over (a_{r_1}^2-{\hbar^2\over 4})^2}
\eeqa
in agreement with formulas (B.4) in \cite{Marino:2004cn}.
 The Seiberg-Witten curve is now given by
 (Table 3 of \cite{Ennes:1999fb})
\beq
S^{Sp(2n)}(x) ={\bar S_0(x)\over (2x)^4}= {1\over (2x)^4\prod_{r=1}^n (x^2-a_r^2)^2}
\label{sfuncsp}
\eeq
Now $\hbar^2 Z^{Sp(2n)}_{0,1}=\ft12 \bar{S}_0(0)^{1\over 2}$ \cite{Ennes:1999fb}.

Formulae (\ref{res},\ref{s02n1},\ref{spn},\ref{zsp}) can then be thought of as a
deformation of the Seiberg and Witten curves to account for
gravitational backgrounds in the gauge theory.

The character corresponding to the case in which matter transforming in the adjoint representation
is added to the lagrangian is obtained  by multiplying (\ref{charor}) by
$\ft12(2-T_m-T_m^{-1})$, see (\ref{traceh}).
This term is telling us that the extra contribution due to this matter is given by the inverse of the
square root of the pure gauge result shifted by $+m$ times with eigenvalues shifted by $\pm m$.
More precisely, the partition functions can then be read from (\ref{res},\ref{s02n1},\ref{spn},\ref{zsp}) by replacing
\beqa
{1\over E_{\alpha \beta}(s)}&\to & {(E_{\alpha \beta}(s)^2-m^2)^{1\over 2} \over  E_{\alpha \beta}(s)}\nn\\
(\phi(s)+x) &\to &\left[(\phi(s)+x)^2-m^2\right]^{-{1\over 2}}
\label{matter}
\eeqa
Although is not obvious at first sight, the partition function after the replacements (\ref{matter})
 does not contain any square roots.
In particular applying (\ref{matter}) to (\ref{sfunc2}) one finds for the massive matter contributions
\beqa
S^{SO(2n)}_m(x) &=& {\prod_{r=1}^n [(x+m)^2-a^2_r][(x-m)^2-a^2_r]\over (2x+ m)^2(2x- m)^2}\nn\\
S^{SO(2n+1)}_m(x) &=& {(x+m)(x-m)\prod_{r=1}^n [(x+m)^2-a^2_r][(x-m)^2-a^2_r]\over (2 x+ m)^2(2x- m)^2}
\label{sfuncancora1}
\eeqa
The $Sp(2n)$ can be treated in an analogous way starting from (\ref{sfuncsp})
\beqa
S^{Sp(2n)}_m(x) &=& (4 x^2-m^2)^2\,\prod_{r=1}^n [(x+m)^2-a^2_r][(x-m)^2-a^2_r]
\label{sfuncancora1sp}
\eeqa

All of these results are in agreement with Table 3 of \cite{Ennes:1999fb}.

The contribution of matter in the fundamental representation, with mass $\pm m_f$, is easily added
introducing the additional term
\beq \delta T_{{\cal M}}=- (V_\phi +
V_\phi^*)(T_{m_f}+T^*_{m_f})
\eeq
in (\ref{tangentspso}). With respect to the discussion
leading to (3.21) in \cite{Bruzzo:2002xf} we have also reflected
the D7-brane with respect to the O3-plane. The character is
modified by a term
\beq \delta \chi=- \sum_{r=1}^n \sum_{s\in
Y_{r}}\left(e^{i(\phi(s)+m_f)}+ e^{i(\phi(s)-m_f)}\right)+{\rm
h.c.}
\eeq
which, in turn, leads to the term
\beq
Z_{m_f}=\prod_{r=1}^n \prod_{s\in Y_{r}}(\phi^2(s)-m_f^2)
\eeq
to
be multiplied in the numerator of the partition functions
(\ref{res},\ref{s02n1},\ref{spn}). This leads to an
additional factor $(x^2-m_f^2)$ in the numerators of the Seiberg-Witten
functions $S(x)$, once again
in agreement with Table 3 of \cite{Ennes:1999fb}.
After this computation was finished Ref.\cite{Shadchin:2004yx} appeared which has an overlap
with the results of this section.

\section*{Acknowledgements}
The authors want to thank R.Flume for collaboration in
the early stage of this work.
R.P. have been partially supported by the Volkswagen foundation
of Germany and he also would like to thank I.N.F.N. for supporting
a visit to the University of Rome II, "Tor Vergata".
This work was supported in part by the EC contract
HPRN-CT-2000-00122, the EC contract HPRN-CT-2000-00148,
the EC contract HPRN-CT-2000-00131,
the MIUR-COFIN contract 2003-023852, the NATO
contract PST.CLG.978785 and the INTAS contracts 03-51-6346 and 00-561.

\begin{appendix}

\section{${\cal N}=(2,2)$ gauge theory}
\label{ad2}
\setcounter{equation}{0}
 We use for the $D=2$ supermultiplets the same notation of their ancestors in $D=4$.
 Supersymmetry transformations follow from field redefinitions
in formulae (2.12) and (2.14) of \cite{Witten:1993yc}
 \footnote{Our conventions: $\delta v_m=i\bar{\epsilon}_{\dot{\alpha}} \bar{\sigma}_{m}^{\dot{\alpha}\alpha}
 \lambda_{\alpha}+   i \epsilon^{\alpha} \sigma_{m,\alpha\dot{\alpha}}
 \bar{\lambda}^{\dot\alpha}$, $\sigma_m=(1,\vec{\sigma}),\bar{\sigma}_m=(1,-\vec{\sigma})$,
 $\lambda_\alpha=\pmatrix{\lambda_-\cr \lambda_+}$, $\bar{\lambda}^{\dot\alpha}=\pmatrix{\bar{\lambda}^-\cr \bar{\lambda}^+}$,
 $\bar{\epsilon}_{\dot\alpha}= (\bar{\epsilon}_- \, \bar{\epsilon}_+)$,
 $\epsilon^{\alpha}= (\epsilon^- \, \epsilon^+)$. In addition we make the replacements:
 $\phi_s\to i \sqrt{2}\,\Phi_s$, $F_s\to \ft{i}{\sqrt{2}}
H_s$ in the chiral multiplets.}.
The four components of the vector field $v_m$ in $D=4$ will be written as
$v_{\pm\pm}\equiv \ft12(v_0\pm v_3)$, $B=\ft12(v_1-i v_2)$ and $\bar{B}=\ft12(v_1+i v_2)$.

Vector Multiplet:
 \beqa
\delta v_{++}&=& i\,  \bar{\epsilon}_{+} \lambda_{+}+i\,
\epsilon_{+} \bar{\lambda}_+ \nn\\
 \delta v_{--}&=& i\,
\bar{\epsilon}_{-} \lambda_{-}+i\, \epsilon_{-} \bar{\lambda}_-\nn\\
\delta B&=& -i\,  \bar{\epsilon}_{+} \lambda_{-}- i\,\epsilon_{-}
\bar{\lambda}_+ \nn\\
 \delta \bar{B}&=& -
i\bar{\epsilon}_{-} \lambda_+-i\,  \epsilon_{+}
\bar{\lambda}_{-} \nn\\
\delta \lambda_+&=&  i \epsilon_{+}\, D-2\,\epsilon_{+}\,F_{-+}+
4\,\epsilon_{-} \nabla_{++} \bar{B}
 \nn\\
 \delta \lambda_-&=&  i\, \epsilon_{-}\,D+2\,\epsilon_{-}\,F_{-+}+
4\, \epsilon_{+} \nabla_{--} B
 \nn\\
 \delta \bar{\lambda}_+&=& - i\, \bar{\epsilon}_{+}\, D-2\,\bar{\epsilon}_{+}\,F_{-+}+
4\, \bar{\epsilon}_{-} \nabla_{++} B
 \nn\\
\delta \bar{\lambda}_-&=& - i \bar{\epsilon}_{-}\, D+ 2\,\bar{\epsilon}_{-}\,F_{-+}+
4\, \bar{\epsilon}_{+} \nabla_{--} \bar{B}\nn\\
\delta D&=& - 2\,\bar{\epsilon}_{+} \nabla_{--} \lambda_{+} -
2\,\bar{\epsilon}_{-} \nabla_{++} \lambda_{-}+2\, \epsilon_{+}
\nabla_{--} \bar{\lambda}_{+}+2\, \epsilon_{-} \nabla_{++}
\bar{\lambda}_{-}
 \label{mult1}
\eeqa with $F_{-+}=[\nabla_{--},\nabla_{++}]=\ft12 v_{03}$.

Chiral multiplet:
\beqa
\delta \Phi_s&=& -i\,  \epsilon_{+}
\Psi_{-,s}+i \epsilon_{-} \Psi_{+,s}  \nn\\
\delta \Psi_{+,s}&=& i \epsilon_{+} H_s-4 \,\bar{\epsilon}_{-}\nabla_{++}\Phi_s \nn\\
\delta \Psi_{-,s}&=& i \epsilon_{-} H_s+4 \,\bar{\epsilon}_{+}\nabla_{--}\Phi_s \nn\\
\delta H_s &=& -4 \,\bar{\epsilon}_{+}\nabla_{--}\Psi_{+,s}-4
\,\bar{\epsilon}_{-}\nabla_{++}\Psi_{-,s}\label{mult2} \eeqa
 Choosing $\bar{\epsilon}_-=\epsilon_-=0$,
$\bar{\epsilon}_+=-\epsilon_+=i\xi$ one finds:
 \beqa
&& Q\, v_{++}=  \bar{\lambda}_+ -\lambda_{+}  \qquad Q\,
(\bar{\lambda}_+ - \lambda_+)= -4 i \,[\nabla_{--},\nabla_{++}] \nn\\
&& Q\, (\bar{\lambda}_+ + \lambda_+) = 2\,D
 \qquad    Q\, D= -2 i\nabla_{--} (\lambda_{+}+\bar{\lambda}_{+})  \nn\\
&&  Q\, B=   \lambda_{-} \qquad  Q\, \lambda_-= -4 i \nabla_{--}B \nn\\
&& Q\, \bar{B}=-\bar{\lambda}_{-} \qquad Q\,
\bar{\lambda}_-=
4 i \nabla_{--} \bar{B}\nn\\
&&Q\, v_{--}= 0
%\label{mult1p}
\nn\\
&& Q\, \Phi_s= -\Psi_{-,s}
 \qquad   Q\, \Psi_{-,s}=  4 i\nabla_{--}\Phi_s \nn\\
&& Q\, \Psi_{+,s}=   H_s \qquad \delta H_s = -4 i
\nabla_{--}\Psi_{+,s} \label{mult2p} \eeqa

Finally, compactifying to zero dimensions and identifying
\beqa
&&\phi=4i v_{--}= -4i\nabla_{--}\nn\\
&&\bar{\phi}= 4 i v_{++}=4 i \nabla_{++}\qquad \quad \eta=4i(\bar{\lambda}_+ -\lambda_+) \nn\\
&& H_\R=2\, D\quad ~~~~~~~~~~~~~~\chi_\R=(\bar{\lambda}_+ -\lambda_+)\nn\\
&& B=B_4  \quad~~~~~~~~~~~~~~ \lambda_-={\cal M}_4 \nn\\
&& \Phi_s=(B_{1,2,3};\omega_{\dot{\alpha}} ) \qquad \Psi_{-,s}=-({\cal M}_{1,2,3};\mu_a)\nn\\
&& H_s=(H_{14,24,34};H_{\dot{a}}) \qquad
\Psi_{+,s}=({\chi}_{14,24,34};\mu_{\dot{a}}) \eeqa one finds
(\ref{brs}).

\section{$SU(2)\times SU(2)$ with one bifundamental}
\label{aquiv}
\setcounter{equation}{0}
For the $\Z_p$ orbifold we find in general a $\prod_r U(N_r)$
gauge theory with bifundamental matter. In particular for the
choice $q_\alpha=0, q_{\hat\alpha}=1$, $N_r=0$ for $r>1$ one finds
again a SYM theory with gauge group $U(N_0)\times U(N_1)$ but now
with a single hypermultiplet.
 Indeed only one of the two combinations, say $a_{r\hat r}+ m$,
survives the orbifold projection and formulae (\ref{replace}) get
modified by the replacements
\beq (a_{r \hat r}+x
)^2-m^2\to (a_{r \hat r}+x) +m
\label{r2}
\eeq
This
results into a single contribution in the numerators of
$S_r(x),S_{\hat r}$ in (\ref{replace}) corresponding
 to the single bifundamental in the resulting quiver.
 This is the case considered in \cite{Ennes:1998ve}. The result (\ref{f4m},\ref{replace}) are indeed
 in perfect agreement with formula (27,28) in \cite{Ennes:1998ve} after (\ref{r2}) is taken
 into account (see also \cite{Gomez-Reino:2003cp}).
 Our formulae can also be tested against the results in \cite{Feichtinger:1999vt} for
the $SU(2)\times SU(2)$ with a single bifundamental matter.
Given that the v.e.v.'s in the two gauge groups are $a_1=-a_2=a$ and $a_{\hat 1}=-a_{\hat 2}=\hat a$,
up to $k=3$ one finds
\beqa {\cal F}&=&\frac{\left ( {{\hat a}}^{2}-{a}^{2}\right )  \, q}{2
\, {a}^{2}} +\frac{\left ( a^2-{\hat a}^2 \right )   \, {\hat q}}{2 \,
{{\hat a}}^{2}} -\frac{\left( {a}^{4}-6 \, {a}^{2} \, {{\hat a}}^{2}+5 \,
{{\hat a}}^{4}\right )  \, {q}^{2}}{64 \, {a}^{6}}
-\frac{\left ( {a}^{2}+{{\hat a}}^{2}\right )  \, q \, {\hat q}}{4 \, {a}^{2} \, {{\hat a}}^{2}}\nn\\
&+&\frac{\left ( {a}^{4}-6 \, {a}^{2} \, {{\hat a}}^{2}+5 \,
{{\hat a}}^{4}\right ) \, {q}^{2} \, {\hat q}}{64 \,{a}^{6} \, {{\hat a}}^{2}}
-\frac{\left ( 5 \, {a}^{4}-6 \, {a}^{2} \, {{\hat a}}^{2}+{{\hat a}}^{4}\right
)  \, {{\hat q}}^{2}}{64 \, {{\hat a}}^{6}} +\frac{\left ( 5 \, {a}^{4}-6 \,
{a}^{2} \, {{\hat a}}^{2}+{{\hat a}}^{4}\right )
\, q \, {{\hat q}}^{2}}{64 \, {a}^{2} \, {{\hat a}}^{6}}\nn\\
&+&\frac{{a}^{2} \, \left ( 9 \, {a}^{4}-14 \, {a}^{2} \,
{{\hat a}}^{2}+5 \, {{\hat a}}^{4}\right ) \, {{\hat q}}^{3}}{\hbox{192} \, {{\hat a}}^{10}}
+\frac{{{\hat a}}^{2} \, \left ( 9 \, {{\hat a}}^{4}-14 \, {{\hat a}}^{2} \, {a}^{2}+5
\, {a}^{4}\right ) \, {q}^{3}}{\hbox{192} \, {a}^{10}}
\nn\\
&+& \epsilon_1^2\left[\frac{\left ( {\hat a}^2-a^2\right ) \, \left (
{a}^{2}-2 \, {{\hat a}}^{2}\right )  \, {q}^{2}}{64 \, {a}^{8}}+
\frac{\left ( a^2-{\hat a}^2\right ) \,
\left ( {{\hat a}}^{2}-2 \, {a}^{2}\right )  \, {{\hat q}}^{2}}{64 \, {{\hat a}}^{8}}\right.\nn\\
&+& \left.\frac{{{\hat a}}^{2} \, \left ( {\hat a}^2-a^2 \right )   \, \left (
16 \, {{\hat a}}^{2}-11 \, {a}^{2}\right ) \, {q}^{3} }{\hbox{192} \,
{a}^{12}} +\frac{{a}^{2} \, \left ( a^2-{\hat a}^2 \right )   \, \left (
16 \, {a}^{2}-11 \, {{\hat a}}^{2}\right ) \, {{\hat q}}^{3} \, }{\hbox{192}
\, {{\hat a}}^{12}}
 \right.\nn\\
&+& \left. \frac{\left ( {a}^{4}-3 \, {a}^{2} \, {{\hat a}}^{2}+2 \,
{{\hat a}}^{4}\right ) \, {q}^{2} \, {\hat q}}{64 \, {a}^{8} \, {{\hat a}}^{2}}
+\frac{\left (  {{\hat a}}^{4}-3 \, {a}^{2} \, {{\hat a}}^{2}+2 {a}^{4}\right )
 \, q \, {{\hat q}}^{2}}{64 \, {a}^{2} \, {{\hat a}}^{8}}\right]+\ldots
\eeqa

\section{$SU(2)\times SU(2)$ with two bifundamentals}
\setcounter{equation}{0}
In this appendix we write the prepotential for the case of a supersymmetric ${\cal N}=2$ SUSY
quiver theory with two massless hypermultiplet in the bifundamental with its
first gravitational correction using the same notations of the previous paragraph.
\beqa {\cal F}&=&-\frac{{\left ( {a}^{2}-{{\hat a}}^{2}\right ) }^{2} \,
q}{2 \, {a}^{2}} -\frac{{\left ( {a}^{2}-{{\hat a}}^{2}\right ) }^{2}
\, {\hat q}}{2 \, {{\hat a}}^{2}} -\frac{{\left ( {a}^{2}-{{\hat a}}^{2}\right )
}^{2} \, \left ( 13 \, {a}^{4}-2 \, {a}^{2} \, {{\hat a}}^{2}
+5 \, {{\hat a}}^{4}\right )  \, {q}^{2}}{64 \, {a}^{6}}\nn\\
&+&\frac{{\left ( {a}^{2}-{{\hat a}}^{2}\right ) }^{2} \, \left (
{a}^{2}+{{\hat a}}^{2}\right ) \, q \, {\hat q}}{2 \, {a}^{2} \,
{{\hat a}}^{2}}-\frac{{\left ( {a}^{2}-{{\hat a}}^{2}\right ) }^{2}
 \, \left ( 5 \, {a}^{4}-2 \, {a}^{2} \, {{\hat a}}^{2}+13 \, {{\hat a}}^{4}\right )  \, {{\hat q}}^{2}}{64 \, {{\hat a}}^{6}}
 \nn\\
 &-&\frac{{\left ( {a}^{2}-{{\hat a}}^{2}\right ) }^{2} \,
\left ( 23 \, {a}^{8}-6 \, {a}^{6} \, {{\hat a}}^{2} +16 \, {a}^{4} \,
{{\hat a}}^{4}-10 \, {a}^{2} \, {{\hat a}}^{6}+9 \, {{\hat a}}^{8}\right )
 \, {q}^{3}}{\hbox{192} \, {a}^{10}}\nn\\
&-&\frac{{\left ( {a}^{2}-{{\hat a}}^{2}\right ) }^{2} \, \left ( 9 \,
{a}^{8}-10 \, {a}^{6} \, {{\hat a}}^{2}
 +16 \, {a}^{4} \, {{\hat a}}^{4}-6 \, {a}^{2} \, {{\hat a}}^{6}+23 \, {{\hat a}}^{8}\right )
 \, {{\hat q}}^{3}}{\hbox{192} \, {{\hat a}}^{10}}\nn\\
&-&\frac{{\left ( {a}^{2}-{{\hat a}}^{2}\right ) }^{2} \, \left (
{a}^{6}+9 \, {a}^{4} \, {{\hat a}}^{2}+11 \, {a}^{2} \, {{\hat a}}^{4}-5 \,
{{\hat a}}^{6}\right )
 \, {q}^{2} \, {\hat q}}{32 \, {a}^{6} \, {{\hat a}}^{2}}\nn\\
& +&\frac{{\left ( {a}^{2}-{{\hat a}}^{2}\right ) }^{2} \, \left ( 5 \,
{a}^{6}-11 \, {a}^{4} \, {{\hat a}}^{2}
 -9 \, {a}^{2} \, {{\hat a}}^{4}-{{\hat a}}^{6}\right )  \, q \, {{\hat q}}^{2}}{32 \, {a}^{2} \, {{\hat a}}^{6}}\nn\\
&+&{\epsilon_1}^{2}\,\left[ \frac{{{\hat a}}^{2} \, {\left (
{a}^{2}-{{\hat a}}^{2}\right ) }^{3} \, {q}^{2}}{32 \, {a}^{8}}
-\frac{{a}^{2} \, {\left ( {a}^{2}-{{\hat a}}^{2}\right ) }^{3}
 \, {{\hat q}}^{2}}{32 \, {{\hat a}}^{8}}-
\frac{\left ( {a}^{4}+6 \, {a}^{2} \, {{\hat a}}^{2} +{{\hat a}}^{4}\right )  \,
q \, {\hat q}}{4 \, {a}^{2} \, {{\hat a}}^{2}} \right.\nn\\&+&\left.
\frac{{\left ( {a}^{2}-{{\hat a}}^{2}\right ) }^{3} \, \left ( 3 \,
{a}^{4}+7 \, {a}^{2} \, {{\hat a}}^{2}-2 \, {{\hat a}}^{4}\right )
 \, {q}^{2} \, {\hat q}}{32 \, {a}^{8} \, {{\hat a}}^{2}}+
\frac{{\left ( {{\hat a}}^{2}-{a}^{2}\right ) }^{3} \, \left ( 3 \,
{{\hat a}}^{4}+7 \, {{\hat a}}^{2} \, {a}^{2}-2 \, {a}^{4}\right )
 \, {{\hat q}}^{2} \, q}{32 \, {{\hat a}}^{8} \, {a}^{2}}
\right.\nn\\
&+& \left. \frac{{{\hat a}}^{2} \, {\left ( {a}^{2}-{{\hat a}}^{2}\right ) }^{3}
\, \left ( 3 \, {a}^{4}-3 \, {a}^{2} \, {{\hat a}}^{2} +8 \,
{{\hat a}}^{4}\right )  \, {q}^{3}}{96 \, {a}^{12}} + \frac{{a}^{2} \,
{\left ( {{\hat a}}^{2}-{a}^{2}\right ) }^{3} \, \left ( 3 \, {{\hat a}}^{4}-3
\, {{\hat a}}^{2} \, {a}^{2} +8 \, {a}^{4}\right )  \, {{\hat q}}^{3}}{96 \,
{{\hat a}}^{12}}\right ]+\ldots  \, \nn \eeqa

\end{appendix}

\end{document}